\documentclass[fleqn,twoside]{article}
\usepackage{espcrc2}

\usepackage{graphicx}
\usepackage[figuresright]{rotating}

\newcommand{\ii}{\'\i}
\newcommand{\AmS}{{\protect\the\textfont2
  A\kern-.1667em\lower.5ex\hbox{M}\kern-.125emS}}

\title{20+ years of Inflation}

\author{J. Garc\ii a-Bellido\address[MCSD]{Theory Division CERN, \\ 
        CH-1211 Gen\`eve 23, Switzerland}%
        \thanks{On leave of absence from Departamento de F\ii sica
Te\'orica, Universidad Aut\'onoma de Madrid, 28049 Cantoblanco, Spain.}}
       
\begin{document}

\begin{abstract}
In this talk I will review the present status of inflationary cosmology
and its emergence as the basic paradigm behind the Standard Cosmological
Model, with parameters determined today at better than 10\% level from
CMB and LSS observations. I will also discuss the recent theoretical
developments on the process of reheating after inflation and model
building based on string theory and D-branes.
\vspace{1pc}
\end{abstract}

\maketitle

\section{INTRODUCTION}

Until recently, inflation~\cite{inflation} was considered a wild idea,
seen with skepticism by many high energy physicists and most
astrophysicists.  Thanks to the recent observations of the cosmic
microwave background (CMB) anisotropies and large scale structure (LSS)
galaxy surveys, it has become widely accepted by the community
and is now the subject of texbooks~\cite{textbooks} and graduate
courses. Furthermore, a month ago, the proponents of the idea were
awarded the prestigious Dirac Medal for it~\cite{Dirac}.

In this short review I will outline the reasons why the inflationary
paradigm has become the backbone of the present Standard Cosmological
Model. It gives a framework in which to pose all the basic cosmological
questions: what is the shape and size of the universe, what is the
matter and energy content of the universe, where did all this matter
come from, what is the fate of the universe, etc. I will describe the
basic predictions that inflation makes, most of which have been
confirmed only recently, while some are imminent, and then explore the
recent theoretical developments on the theory of reheating after
inflation and cosmological particle production, which might allow us to
answer some of the above questions in the future.

Although the simplest slow-roll inflation model is consistent with the
host of high precision cosmological observations of the last few years,
we still do not know what the true nature of the inflaton is: although
there are many possible realizations, there is no unique particle
physics model of inflation. Furthermore, we even ignore the energy scale
at which this extraordinary phenomenon occurred in the early universe;
it could be associated with a GUT theory or even with the EW theory, at
much lower energies.

In the last section, I will describe a recent avenue of proposals for
the origin of inflation based on string theory and, in particular, on
extended multidimensional objects called D-branes.  Whether these ideas
will be the seed for the final theory of inflation nobody knows, but it
opens up many new possibilities, and even a new language for the early
universe, that makes them certainly worthwhile exploring.

\section{BASIC PREDICTIONS OF INFLATION}

Inflation is an extremely simple idea based on the early universe
dominance of a vacuum energy density associated with a hypothetical
scalar field called the inflaton. Its nature is not known: whether it is
a fundamental scalar field or a composite one, or something else
altogether. However, one can always use an effective description in
terms of a scalar field with an effective potential driving the
quasi-exponential expansion of the universe. This basic scenario gives
several detailed fundamental predictions: a flat universe with nearly
scale-invariant adiabatic density perturbations with Gaussian initial
conditions.

\subsection{A flat and homogeneous background}

Inflation explains why our local patch of the universe is spatially
flat, i.e.  Euclidean. Such a (geometrical) property is unstable under
the evolution equations of the Big Bang theory in the presence of
ordinary matter and radiation~\cite{textbooks}, making it very unlikely
today, unless some new mechanism in the early universe, prior to the
radiation era (at least as far back as primordial nucleosynthesis)
prepared the universe with such a peculiar initial condition. That is
precisely what inflation does, and very efficiently in fact, by
providing an approximately constant energy density that induces a
tremendous expansion of the universe. Thus, an initially curved
three-space will quickly become locally indistinguishable from a
``flat'' hypersurface.\footnote{Note the abuse of language here. The
universe is obviously {\em not} two-dimensionally 'flat' but 3D
Euclidean.}  Moreover, this same mechanism explains why we see no
ripples, i.e. no large inhomogeneities, in the space-time fabric,
e.g. as large anisotropies in the temperature field of the cosmic
microwave background when we look in different directions.  The
expansion during inflation erases any prior inhomogeneities.

These two are very robust predictions of inflation, and constitute the
0-th order, i.e. the space-time background, in a linear expansion in
perturbation theory.  They have been confirmed to high precision by the
detailed observations of the CMB, first by COBE (1992)~\cite{COBE} for
the large scale homogeneity, to one part in $10^5$, and recently by
BOOMERanG~\cite{deBernardis:2000gy} and MAXIMA~\cite{Hanany:2000qf}, for
the spatial flatness, to better than 10\%.

\subsection{Cosmological perturbations} 

Inflation also predicts that on top of this homogeneous and flat
space-time background, there should be a whole spectrum of cosmological
perturbations, both scalar (density perturbations) and tensor
(gravitational waves). These arise as quantum fluctuations of the metric
and the scalar field during inflation, and are responsible for a scale
invariant spectrum of temperature and polarization fluctuations in the
CMB, as well as for a stochastic background of gravitational waves. The
temperature fluctuations were first discovered by COBE and later
confirmed by a host of ground and balloon-borne experiments, while the
polarization anisotropies have only recently been discovered by the CMB
experiment DASI~\cite{pol}. Both observations seem to agree with a
nearly scale invariant spectrum of perturbations. It is expected that
the stochastic background of gravitational waves produced during
inflation could be detected with the next generation of gravitational
waves interferometers (e.g. LISA), or indirectly by measuring the power
spectra of polarization anisotropies in the CMB by the future Planck
satellite~\cite{Planck}.

Inflation makes very specific predictions as to the nature of the scalar
perturbations. In the case of a single field evolving during inflation,
the perturbations are predicted to be adiabatic, i.e. all components of
the matter and radiation fluid should have equal density contrasts, due
to their common origin. As the plasma (mainly baryons) falls in the
potential wells of the metric fluctuations, it starts a series of
acoustic compressions and rarefactions due to the opposing forces of
gravitational collapse and radiation pressure.  Adiabatic fluctuations
give a very concrete prediction for the position and height of the
acoustic peaks induced in the angular power spectrum of temperature and
polarization anisotropies. This has been confirmed to better than 1\% by
the recent observations, and constitutes one of the most important
signatures in favor of inflation, ruling out a hypothetically large
contribution from active perturbations like those produced by cosmic
strings or other topological defects.

Furthermore, the quantum origin of metric fluctuations generated during
inflation allows one to make a strong prediction on the statistics of
those perturbations: inflation stretches the vacuum state fluctuations
to cosmological scales, and gives rise to a Gaussian random field, and
thus metric fluctuations are in principle characterized solely by their
two-point correlation function. Deviations from Gaussianty would
indicate a different origin of fluctuations, e.g. from cosmic defects.
Recent observations by BOOMERanG in the CMB and by gravitational lensing
of LSS indicate that the non-Gaussian component of the temperature
fluctuations and the matter distribution on large scales is strongly
constrained, and consistent with foregrounds (in the case of CMB) and
with non-linear gravitational collapse (in the case of LSS).

\begin{figure}[htb]
\vspace{-3cm}\hspace{-6mm}
\includegraphics[angle=0,width=8cm]{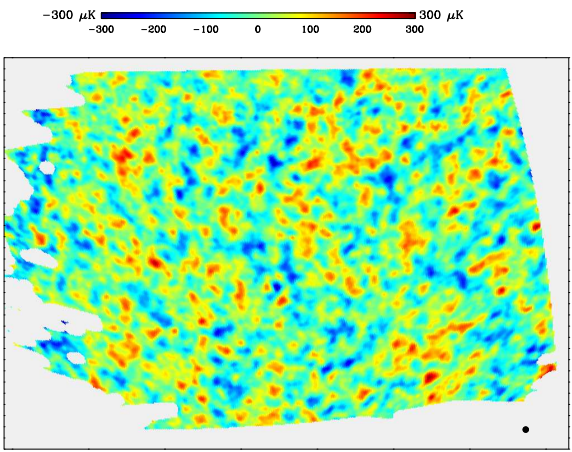}
\caption{The CMB temperature fluctuations as seen by BOOMERanG. The
1 degree features corresponding to the first acoustic peak are clearly
distinguishable. From Ref.~\cite{deBernardis:2000gy}.}
\label{fig:BoomerangMap}
\end{figure}

Of course, in order to really confirm the idea of inflation one needs to
find cosmological observables that will allow us to correlate the scalar
and the tensor metric fluctuations with one another, since they both
arise from the same inflaton field fluctuations. This is a daunting
task, given that we ignore the absolute scale of inflation, and thus the
amplitude of tensor fluctuations (only sensitive to the total energy
density). The smoking gun could be the observation of a stochastic
background of gravitational waves by the future gravitational wave
interferometers and the subsequent confirmation by detection of the curl
component of the polarization anisotropies of the CMB. Although the
gradient component has recently been detected by DASI, we may have to
wait for Planck for the detection of the curl component.

\section{RECENT COSMOLOGICAL OBSERVATIONS}

Cosmology has become in the last few years a phenomenological science,
where the basic theory (based on the hot Big Bang model after inflation)
is being confronted with a host of cosmological observations, from the
microwave background to the large scale distribution of matter, from the
determination of light element abundances to the detection of distant
supernovae that reflect the acceleration of the universe, etc. I will
briefly review here the recent observations that have been used to
define a consistent cosmological standard model.

\subsection{Cosmic Microwave Background}

The most important cosmological phenomenon from which one can extract
essentially all cosmological parameters is the microwave background and,
in particular, the last scattering surface temperature and polarization
anisotropies, see Fig.~\ref{fig:BoomerangMap}. Since they were
discovered by COBE in 1992, the temperature anisotropies have lived to their
promise. They allow us to determine a whole set of both background (0-th
order) and perturbation (1st-order) parameters -- the geometry, topology
and evolution of space-time, its matter and energy content, as well as
the amplitude and tilt of the scalar and tensor fluctuation power
spectra -- in some cases to better than 10\% accuracy.

\begin{figure}[htb]
\vspace{-10pt}\hspace{-20pt}
\includegraphics[angle=0,width=20pc]{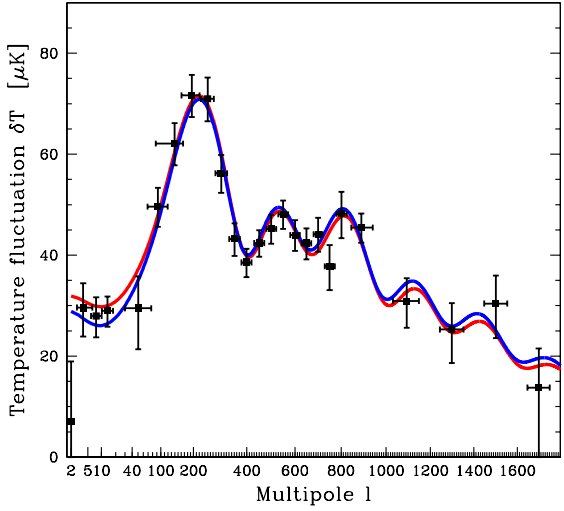}
\vspace{-40pt}
\caption{The angular power spectrum of CMB temperature
anisotropies. Data includes the recent measurements of BOOMERanG,
MAXIMA, DASI, VSA and CBI, as well as a binning of previous measurements,
inluding COBE's. The lines correspond to the theoretical expectation for
the $\Lambda$CDM concordance model of \cite{Wang:2001gy} (red) and
\cite{Efstathiou:2002} (blue). From Ref.~\cite{Tegmark:2002cy}.}
\label{fig:ClTegmark}
\end{figure}

At present, the forerunners of CMB experiments are two balloons --
BOOMERanG~\cite{Boom} and MAXIMA~\cite{Maxima} -- and three ground based
interferometers -- DASI~\cite{DASI}, VSA~\cite{VSA} and
CBI~\cite{CBI}. Together they have allowed cosmologists to determine the
angular power spectrum of temperature fluctuations down to multipoles
1000 and 3000, respectively, and therefore provided a measurement of the
positions and heigths of at least 3 to 7 acoustic peaks, see
Fig.~\ref{fig:ClTegmark}. A combined analysis of the different CMB
experiments yields convincing evidence that the universe is flat, with
$|\Omega_K| = |1-\Omega_{\rm tot}| < 0.05$ at 95\% c.l.; full of dark
energy, $\Omega_\Lambda = 0.66\pm0.06$, and dark matter, $\Omega_m =
0.33\pm0.07$, with about 5\% of baryons, $\Omega_b = 0.05\pm0.01$; and
expanding at a rate $H_0 = 68\pm7$ km/s/Mpc, all values given with
$1\sigma$ errors, see Table~\ref{table}. The spectrum of primordial
perturbations that gave rise to the observed CMB anisotropies is nearly
scale-invariant, $n_s = 1.02\pm0.06$, adiabatic and Gaussian
distributed. This set of parameters already constitutes the basis for a
truly Standard Model of Cosmology, based on the Big Bang theory and the
inflationary paradigm.  Note that both the baryon content and the rate
of expansion determinations with CMB data alone are in excellent
agreement with direct determinations from BBN light element
abundances~\cite{O'Meara:2000dh} and HST
Cepheids~\cite{Freedman:2000cf}, respectively.

\begin{sidewaystable}
\caption{Estimates of the cosmological parameters that characterize a minimal adiabatic inflation-based model. From Ref.~\cite{CBI}.}
\label{table}
\newcommand{\m}{\hphantom{$-$}}
\newcommand{\cc}[1]{\multicolumn{1}{c}{#1}}
\renewcommand{\arraystretch}{1.2} 
\begin{tabular*}{\textheight}{@{\extracolsep{\fill}}lllllllllllll}
\hline\hline
& Priors & \cc{$\Omega_{tot}$} & \cc{$n_s$} & \cc{$\Omega_b\,h^2$} &
\cc{$\Omega_{cdm}\,h^2$} & \cc{$\Omega_\Lambda$} & \cc{$\Omega_m$} & 
\cc{$\Omega_b$} & \cc{$h$} & \cc{Age} & \cc{$\tau_c$} &  \\
\hline
 &  CMB  & $1.05^{0.05}_{0.05}$ & $1.02^{0.06}_{0.07}$ & $0.023^{0.003}_{0.003}$ & $0.13^{0.03}_{0.02}$ & $0.54^{0.12}_{0.13}$ & $0.52^{0.15}_{0.15}$ & $0.080^{0.023}_{0.023}$ & $0.55^{0.09}_{0.09}$ & $15.0^{1.1}_{1.1}$ & $0.16^{0.18}_{0.13}$ & \\[2pt]
 &  CMB+LSS  & $1.03^{0.03}_{0.04}$ & $1.00^{0.06}_{0.06}$ & $0.023^{0.003}_{0.003}$ & $0.12^{0.02}_{0.02}$ & $0.61^{0.09}_{0.10}$ & $0.42^{0.12}_{0.12}$ & $0.067^{0.018}_{0.018}$ & $0.60^{0.09}_{0.09}$ & $14.7^{1.2}_{1.2}$ & $0.09^{0.12}_{0.07}$ & \\[2pt]
 &  CMB+LSS+SN  & $1.00^{0.03}_{0.02}$ & $1.03^{0.06}_{0.06}$ & $0.024^{0.003}_{0.003}$ & $0.12^{0.02}_{0.02}$ & $0.69^{0.04}_{0.06}$ & $0.32^{0.06}_{0.06}$ & $0.052^{0.011}_{0.011}$ & $0.68^{0.06}_{0.06}$ & $13.8^{0.9}_{0.9}$ & $0.13^{0.14}_{0.10}$ & \\[2pt]
 &  CMB+LSS+SN+HST  & $1.00^{0.02}_{0.02}$ & $1.04^{0.05}_{0.06}$ & $0.024^{0.002}_{0.003}$ & $0.12^{0.01}_{0.01}$ & $0.70^{0.02}_{0.03}$ & $0.30^{0.02}_{0.02}$ & $0.049^{0.004}_{0.004}$ & $0.69^{0.02}_{0.02}$ & $13.6^{0.2}_{0.2}$ & $0.13^{0.13}_{0.10}$ & \\
\hline
\end{tabular*}\\[2pt]
The age of the Universe is in Gyr, and the rate of expansion in units
of 100 km/s/Mpc. All values quoted with $1\sigma$ errors.
\end{sidewaystable}

In the near future, a new satellite experiment, the Microwave Anisotropy
Probe (MAP)~\cite{MAP}, will provide a full-sky map of temperature (and
possibly also polarization) anisotropies and determine the first 2000
multipoles with unprecedented accuracy. When combined with LSS and SN
measurements, it promises to allow the determination of most cosmological
parameters with errors down to the few\% level.

\begin{figure}[htb]
\vspace{5pt}\hspace{2pt}
\includegraphics[angle=0,width=17pc]{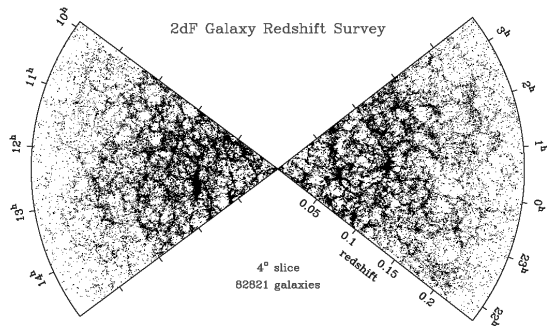}
\vspace{-10pt}
\caption{The galaxy distribution of the 2dFGRS, drawn from a total of
221283 galaxies (15 May 2002) in slices $4^\circ$ thick, centered at
declination $-2.5^\circ$ in the NGP and $-27.5^\circ$ in the SGP.
From Ref.~\cite{Peacock:2002}.}
\label{fig:2dFslice}
\end{figure}

Moreover, with the recent detection of microwave background polarization
anisotropies by DASI~\cite{pol}, confirming the basic paradigm behind
the Cosmological Standard Model, a new window opens which will allow yet
a better determination of cosmological parameters, thanks to the very
sensitive (0.1$\mu$K) and high resolution (4 arcmin) future satellite
experiment Planck~\cite{Planck}. In principle, Planck should be able to
detect not only the gradient component of the CMB polarization, but also
the curl component, if the scale of inflation is high enough. In that
case, there might be a chance to really test inflation through
cross-checks between the scalar and tensor spectra of fluctuations,
which are predicted to arise from the same inflaton potential.

The observed positions of the acoustic peaks of the CMB anisotropies
strongly favor purely adiabatic density perturbations, as arise in the
simplest single-scalar-field models of inflation. These models also
predict a nearly Gaussian spectrum of primordial perturbations. A small
degree of non-gaussianity may arise from self-coupling of the inflaton
field (although it is expected to be very tiny, given the observed small
amplitude of fluctuations), or from two-field models of inflation. Since
the CMB temperature fluctuations probe directly primordial density
perturbations, non-gaussianity in the density field should lead to a
corresponding non-gaussianity in the temperature maps. However, recent
searches for non-Gaussian signatures in the CMB have only given
stringent upper limits, see Ref.~\cite{Polenta:2002}.

\begin{figure}[htb]
\vspace{0pt}\hspace{-13pt}
\includegraphics[angle=0,width=19pc]{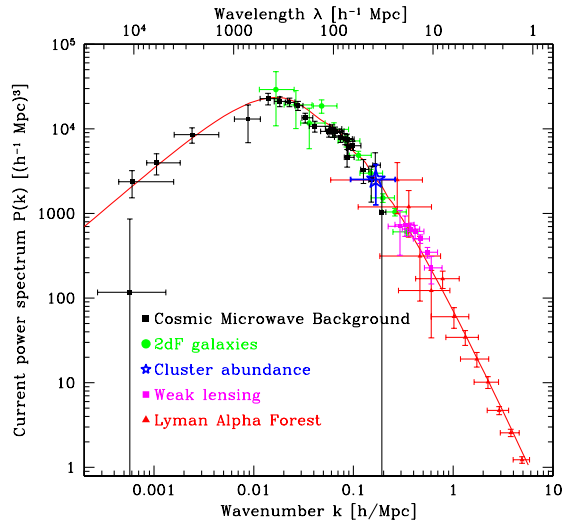}
\vspace{-40pt}
\caption{The matter power spectrum. It includes data from the CMB, the
matter distribution as seen by 2dFGRS, the cluster abundance, weak
gravitational lensing, and Lyman-$\alpha$ forest data. From
Ref.~\cite{Tegmark:2002cy}.}
\label{fig:PkTegmark}
\end{figure}

One of the most interesting aspects of the present progress in
cosmological observations is that they are beginning to probe the same
parameters or the same features at different time scales in the
evolution of the universe. We have already mentioned the determination
of the baryon content, from BBN (light element abundances) and from the
CMB (acoustic peaks), corresponding to totally different physics and yet
giving essentially the same value within errors. Another example is the
high resolution images of the CMB anisotropies by CBI~\cite{CBI}, which
constitute the first direct detection of the seeds of clusters of
galaxies, the largest gravitationally bound systems in our present
universe. In the near future we will be able to identify and put into
one-to-one correspondence tiny lumps in the CMB with actual clusters
today.

\subsection{Large Scale Structure}

The last decade has seen a tremendous progress in the determination
of the distribution of matter up to very large scales. The present
forerunners are the 2dF Galaxy Redshift Survey~\cite{2dFGRS} and the 
Sloan Digital Sky Survey (SDSS)~\cite{SDSS}. These deep surveys aim
at $10^6$ galaxies and reach redshifts of order 1 for galaxies
and order 5 for quasars. They cover a wide fraction of the sky and
therefore can be used as excellent statistical probes of large scale
structure~\cite{Peacock:2002,Percival:2002}, see Fig.~\ref{fig:2dFslice}.

The main output of these galaxy surveys is the two-point (and higher)
spatial correlation functions of the matter distribution or,
equivalently, the power spectrum in momentum space. Given a concrete
type of matter, e.g. adiabatic vs. isocurvature, cold vs. hot, etc., the
theory of linear (and non-linear) gravitational collapse gives a very
definite prediction for the measured power spectrum, which can then be
compared with observations. This quantity is very sensitive to various
cosmological parameters, mainly the dark matter content and the baryonic
ratio to dark matter, as well as the universal rate of expansion; on the
other hand, it is mostly insensitive to the cosmological constant since
the latter has only recently (after redshift $z\sim1$) started to become
important for the evolution of the universe, while galaxies and clusters
had already formed by then.

In Fig.~\ref{fig:PkTegmark} we have plotted the measured matter power
spectrum with LSS data from the 2dFGRS, plus CMB, weak gravitational
lensing and Lyman-$\alpha$ forest data. Together they allow us to
determine the power spectrum with better than 10\% accuracy for $k>0.02
\ h$ Mpc$^{-1}$, which is well fitted by a flat CDM model with
$\Omega_m\,h = 0.20 \pm 0.03$, and a baryon fraction of $\Omega_b /
\Omega_m = 0.15 \pm 0.06$, which together with the HST results
give values of the parameters that are compatible with those obtained
with the CMB.

\begin{figure}[htb]
\includegraphics[angle=0,width=17pc]{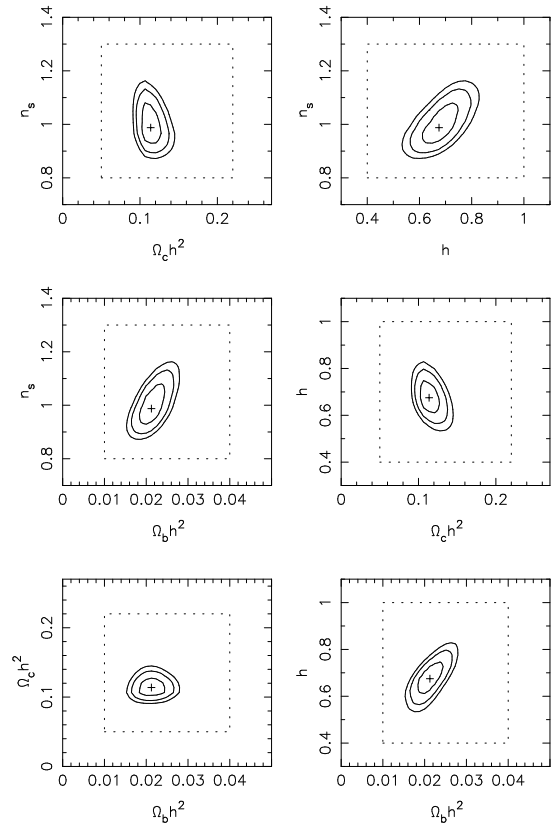}
\vspace{-10pt}
\caption{Likelihood contours for a fit to current CMB data plus the
2dFGRS power spectrum, assuming a flat scalar-dominated model. Note the
10\% precision already achieved by cosmological observations. The values
of the Hubble parameter and baryon density are in very good agreement
with direct estimates. From Ref.~\cite{Peacock:2002}.}
\label{fig:Peacock}
\end{figure}

However, the greatest power is attained when combining CMB with LSS, as
can be seen in Table~\ref{table} and Fig.~\ref{fig:Peacock}. It is very
reassuring to note that present parameter determination is robust as we
progress from weak priors to the full cosmological information
available, a situation very different from just a decade ago, where the
errors were mostly systematic and parameters could only be determined
with an order-of-magnitude error. In the very near future such errors
will drop again to the 1\% level, making Cosmology a mature science,
with many independent observations confirming and further constraining
previous measurements of the basic parameters.

An example of such progress appears in the analysis of non-Gaussian
signatures in the primordial spectrum of density perturbations. The
tremendous increase in data due to 2dFGRS and SDSS has allowed
cosmologists to probe the statistics of the matter distribution on very
large scales and infer from it that of the primordial spectrum.
Recently, both groups have reported non-Gaussian signatures (in
particular the first two higher moments, skewness and kurtosis), that
are consistent with gravitational collapse of structure that was
originally Gaussianly distributed~\cite{Verde:2000vr,Szapudi:2001mh}. 

Moreover, weak gravitational lensing also allows an independent
determination of the three-point shear correlation function, and there
has recently been a claim of detection of non-Gaussian signatures in the
VIRMOS-DESCART lensing survey~\cite{Bernardeau:2002}, which is also
consistent with theoretical expectations of gravitational collapse of
Gaussianly distributed initial perturbations.

The recent precise catalogs of the large scale distribution of matter
allows us to determine not only the (collapsing) cold dark matter
content, but also put constraints on the (diffusing) hot dark matter,
since it would erase all structure below a scale that depends on the
free streaming length of the hot dark matter particle. In the case of
relic neutrinos we have extra information because we know precisely
their present energy density, given that neutrinos decoupled when the
universe had a temperature around 0.8 MeV and cooled down ever
since. Their number density today is around 100 neutrinos/cm$^3$. If
neutrinos have a significant mass (above $10^{-3}$ eV, as observations
of neutrino oscillations by SuperKamiokande~\cite{SK} and Sudbury
Neutrino Observatory~\cite{SNO} seem to indicate), then the relic
background of neutrinos is non-relativistic today and could contribute a
large fraction of the critical density, $\Omega_\nu = m_\nu/92\,h^2\!
{\rm eV}\geq 0.001$, see Ref.~\cite{Hu:1997mj}. Using observations of 
the Lyman-$\alpha$ forest in absorption spectra of quasars, due to a
distribution of intervening clouds, a limit on the absolute mass of all
species of neutrinos can be obtained~\cite{Croft:1999mm}. Recently, the
2dFGRS team~\cite{Elgaroy:2002bi} have derived a bound on the allowed
amount of hot dark matter, $\Omega_\nu < 0.13\,\Omega_m < 0.05$ (95\%
c.l.), which translates into an upper limit on the total neutrino mass,
$m_{\nu,tot} < 1.8$ eV, for values of $\Omega_m$ and the Hubble constant
in agreement with CMB and SN observations. This bound improves several
orders of magnitude on the direct experimental limit on the muon and tau
neutrino masses, and is comparable to present experimental bounds on
the electron neutrino mass~\cite{PDG}.

\subsection{Cosmological constant and rate of expansion}

Observations of high redshift supernovae by two independent groups, the
Supernova Cosmology Project~\cite{SNCP}, and the High Redshift Supernova
Team~\cite{HRST}, give strong evidence that the universe is
accelerating, instead of decelerating, today. Although a cosmological
constant is the natural suspect for such a ``crime'', its tiny non-zero
value makes theoretical physicists
uneasy~\cite{Weinberg,Carroll,Peebles}.

A compromise could be found -- although to my taste it is rather ugly --
by setting the fundamental cosmological constant to zero, by some yet
unknown principle possibly related with quantum gravity, and allow a
super-weakly-coupled homogeneous scalar field to evolve down an almost
flat potential. Such a field would induce an effective cosmological
constant that could in principle account for the present observations.
The way to distinguish it from a true cosmological constant would be
through its equation of state, since such a type of smooth background is
a perfect fluid but does not satisfy $p=-\rho$ exactly, and thus
$w=p/\rho$ also changes with time. There is a proposal for a satellite
called the Supernova / Acceleration Probe (SNAP)~\cite{SNAP} that will
be able to measure the light curves of type Ia supernovae up to redshift
$z\sim2$, thus determining both $\Omega_X$ and $w_X$ with reasonable
accuracy, where $X$ stands for this hypothetical scalar field. For the
moment there are only upper bounds, $w_X < -0.6$ (95\% c.l.)~\cite{PTW},
consistent with a true cosmological constant, but the SNAP project
claims it could determine $\Omega_X$ and $w_X$ with 5\% precision.

Fortunately, the SN measurements of the acceleration of the universe
give a linear combination of cosmological parameters that is almost
orthogonal, in the plane ($\Omega_m,\ \Omega_\Lambda$), to that of the
curvature of the universe ($1-\Omega_K = \Omega_m + \Omega_\Lambda$) by
CMB measurements and the matter content by LSS data. Therefore, by
combining the information from SNe with that of the CMB and LSS, one can
significantly reduce the errors in both $\Omega_m$ and $\Omega_\Lambda$,
see Table~\ref{table}. It also allows an independent determination of
the rate of expansion of the universe that is perfectly compatible with
the HST data~\cite{Freedman:2000cf}.  This is reflected on the fact that
adding the latter as prior does not affect significantly the mean value
of most cosmological parameters, only the error bars, and can be taken
as an indication that we are indeed on the right track: the Standard
Cosmological Model is essentially correct, we just have to improve the
measurements and reduce the error bars.

\section{OTHER PREDICTIONS OF INFLATION}

Inflation not only predicts the large scale flatness and homogeneity of
the universe and a scale invariant spectrum of metric perturbations.  It
also implies that all the matter and energy we see in the universe today
came from the approximately constant energy density during inflation.
The process that converts this energy density into thermal radiation is
called reheating of the universe and occurs as the inflaton decay
produces particles that rescatter and finally thermalise. In the last
decade there has been a tremendous theoretical outburst of activity in
this field, thanks to the seminal paper of Kofman, Linde and
Starobinsky~\cite{KLS}, proposing that reheating could have been
preceded by a (short) period of explosive particle production that is
extremely non linear, non perturbative and very far from equilibrium.
It is this so-called {\em preheating} process which is probably
responsible for the huge comoving entropy present in our universe. This
period is so violent that it may have produced a large stochastic
background of gravitational waves~\cite{GW}, which may be detected by
future gravitational wave observatories. Such a signature would open up
a new window into the very early universe, independently of the late CMB
and LSS observations.

\subsection{Preheating after inflation}

Preheating occurs due to the coherent oscillations of the inflaton at
the end of inflation, which induces, through parametric resonance, a
short burst of particle production. It is a non-perturbative process
very similar to the Schwinger pair production mechanism in quantum
electrodynamics. The process of preheating through parametric resonance
has been studied in essentially all models of inflation~\cite{KLS},
including hybrid inflation~\cite{hybrid}, and seen to be very
generic~\cite{GBL}. Rescattering occurs soon after preheating and
thermalization eventually involves all species which are coupled to one
another, even if they are not coupled to the inflaton~\cite{FK}.
Nowadays, the full study of this non-perturbative process can only be
done with real time numerical lattice simulations~\cite{TK}, performed
with a C++ program called LATTICEEASY~\cite{FT}.

Recently, the process of reheating has been studied in the context of
theories with spontaneous symmetry breaking~\cite{GBKLT}, and seen to
involve strong particle production from the tachyonic growth of the
symmetry breaking (Higgs) field~\cite{Garcia-Bellido:2002aj}, and not
from the coherent oscillations of the inflaton field. This new process
has been denominated as {\em tachyonic preheating}~\cite{GBKLT}, and
shown to be very efficient in the production of other particles like
bosons or fermions that coupled to the Higgs
field~\cite{Garcia-Bellido:2001cb,CPR}.

\subsection{Baryo- and lepto-genesis}

This non-perturbative and strongly out of equilibrium stage of the
universe could be responsible for one of the remaining mysteries of the
universe: the origin of the observed matter-antimatter
asymmetry~\cite{textbooks}, characterised by the conserved ratio of 1
baryon for every $10^9$ photons. For the baryon asymmetry of the
universe to occur it is necessary (but not sufficient) that the three
Sakharov conditions be satisfied: B-violation, C- and CP-violation, and
out of equilibrium evolution. 

A possible scenario is that of the GUT baryogenesis occurring at the end
of inflation through the far from equilibrium stage associated with
preheating~\cite{Kolb:1996jt}, or even from couplings of the inflaton to
R-handed fermions that decay into SM Higgs and leptons, giving rise to a
lepton asymmetry~\cite{Giudice:1999fb}, which later gets converted into
a baryon asymmetry through sphaleron transitions at the electroweak
scale, a process known as leptogenesis. Another possibility is that the
process of leptogenesis occurs at the end of a hybrid model of
inflation, where the couplings of the GUT Higgs field induces strong
R-handed fermion production again giving rise to a lepton
asymmetry~\cite{Garcia-Bellido:2001cb}.

Moreover, preheating at the electroweak (EW) scale could be sufficiently
strongly out of equilibrium to render viable the process of electroweak
baryogenesis~\cite{GBKS,KT}, as long as a new source of CP-violation is
assumed at the TeV; see also Ref.~\cite{RSC}. The only assumption
needed for preheating to occur at the EW scale is that a short secondary
stage of inflation occurs at low energies, as in thermal
inflation~\cite{LS}, or through some finetuning of
parameters~\cite{review}. The CMB anisotropies need not arise from the
same stage.

\section{MODEL BUILDING}

Given the tremendous success that inflationary cosmology is having in
explaining the present CMB and LSS observations, it would be desirable
that a concrete particle physics model were directly responsible for it.
In that case, one could ask many more questions about the model, like
couplings to other fields and the decay rate of the inflaton, which
would be directly related to the final reheating temperature and the
process of preheating. For the moment the CMB observations do not give
us much clues, although in the near future, with the MAP satellite, and
certainly with Planck, knowledge of the spectral index of density
perturbations, and a possible presence of a tensor (gravitational wave)
component in the polarization anisotropies, will constraint tremendously
the different models already proposed. It is therefore the challenge of
theoretical cosmologists to come up now with new proposals for models of
inflation that can withstand the expected level of accuracy from
future observations.

\subsection{hybrid inflation}

Perhaps the best motivated model of inflation, from the point of view of
particle physics, is hybrid inflation~\cite{hybrid}. It is the simplest
posible extension of a symmetry breaking (Higgs) field coupled to a
singlet scalar that evolves along its slow-roll potential driving
inflation with the false vacuum energy of the Higgs field. Inflation
ends as the inflaton slow-rolls below a critical value that produces a
change of sign of the Higgs' effective mass squared, from positive to
negative, triggering spontaneous symmetry breaking through spinodal
instability~\cite{GBL,ABC,CPR,Garcia-Bellido:2002aj}. In most models the
end of inflation occurs in less than one e-fold, and the false vacuum
energy decays into radiation through tachyonic preheating~\cite{GBKLT}.
In others there are still a few e-folds after the bifurcation point, and
a large peak appears in the power spectrum of density
fluctuations~\cite{Randall:1995dj}, which could be responsible for
primordial black hole formation~\cite{GBLW}.  Preheating in hybrid
inflation was first studied in Ref.~\cite{GBL}, where both types of
models were considered and shown to be very different in the process of
reheating the universe. This may help distinguish them in the future.

Hybrid inflation has a natural realization in the context of
supersymmetric theories, both in D-term and F-term type
models~\cite{review}.  It has the advantage that the values of the
fields during inflation are well below the Planck scale, so that
supergravity corrections can be safely ignored~\cite{Linde:1997sj}.
Moreover, in susy hybrid inflation it is possible to compute the
radiative corrections and show that inflation occurs along the 1-loop
Coleman-Weinberg effective potential~\cite{DSS}. Preheating in a
concrete susy hybrid inflation model was studied in Ref.~\cite{BGK}.

\subsection{String theory and D-branes}

String theory as a theory of everything and, in particular, as a theory
of quantum gravity~\cite{Polchinski}, is a natural framework in which to
search for a fundamental theory of inflation. After many years such a
search had not been very successful, because the low energy effective
theory of strings does not contain any natural candidate for an inflaton
field with a sufficiently flat potential to ensure agreement with CMB
observations. Only recently, with the advent of D-branes as
non-perturbative extended objects in string theory, has there been some
progress towards a theory of inflation based on string
theory~\cite{braneinflation}. The most striking characteristic of brane
inflation is the geometrical interpretation of the inflaton: it is not a
field in the spectrum of strings but the coordinate distance between two
D-branes or two orientifold planes in the internal (compactified)
space. The scalar potential driving inflation is exactly computable from
the 1-loop exchange of open string modes or, equivalently, from the tree
level exchange of closed string modes like the graviton, dilaton, etc.
In the limit of large distances between the branes (in units of the
string length) the potential reduces to the ``Newtonian'' potential of a
supergravity theory due to the exchange of the massless modes of the
theory in the compact dimensions,
\begin{equation}\label{pot}
V(y) = V_0 - \beta/y^{d_\perp-2}
\end{equation}
where $V_0$ is the tree level exchange, the false vacuum energy driving
inflation, $y$ is the distance between branes/orientifolds, which plays
the role of the inflaton, $d_\perp$ is the number of transverse
dimensions to the brane in the compactified space, and $\beta$ is a
constant that depends on the string coupling, the number of massless
fields exchanged and the orientation of the branes. In early models of
brane inflation it was necessary to finetune the coupling $\beta$ in
order to have enough e-folds of inflation~\cite{braneinflation}. Last
year we proposed a solution~\cite{GBRZ} by tilting slightly the relative
orientation of the branes through a small angle $\theta$ related to
supersymmetry breaking in the bulk. In that case, the coupling $\beta$
is naturally small and a successful inflation is ensured. The potential
is attractive and the two branes approach eachother. When the branes are
close enough, compared with the string scale, there appears a tachyon in
the spectrum of the strings, whose effective potential develops a
minimum and a true vacuum. Inflation ends when the inflaton changes the
sign of the tachyon mass squared and triggers symmetry breaking.

\begin{figure}[htb]
\vspace{5pt}\hspace{1cm}
\includegraphics[angle=0,width=5cm]{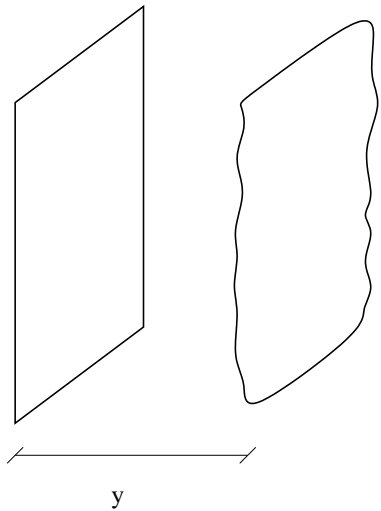}
\vspace{-10pt}
\caption{The inflaton field is interpreted as the distance $y$ between
the two branes. Quantum fluctuations of this field will give rise upon
collision to density perturbations on comoving hypersurfaces. These
fluctuations will be later observed as temperature anisotropies in
the microwave background. From Ref.~\cite{GBRZ}.}
\label{fluctuations}
\end{figure}

\begin{table*}[htb]
\caption{Correspondences between ordinary and brane inflation.}
\label{table2}
\newcommand{\m}{\hphantom{$-$}}
\newcommand{\cc}[1]{\multicolumn{1}{c}{#1}}
\renewcommand{\tabcolsep}{2pc} 
\renewcommand{\arraystretch}{1.2} 
\begin{tabular}{@{}ll}
\hline\hline
Quantum Field Theory  & D-Brane/String Theory \\
\hline
Scalar field $\phi$  &  Distance between branes $\ y$ \\
Potential $V(\phi)$  &  Tree-level + 1-loop exchange of string 
modes~(\ref{pot}) \\
\hline
Slow-roll parameters  &  Slow-roll parameters  \\[2mm]
$\epsilon = {\displaystyle
{M_{\rm P}^2\over2}\left({V'\over V}\right)^2}$  &  
$\epsilon = {\displaystyle
{3v^2\over4}}$ \hspace{2.65cm} {\rm relative velocity 
of the branes}\\[1mm]
$\eta = {\displaystyle M_{\rm P}^2\,{V''\over V}}$  &  
$\eta = {\displaystyle
- {d_\perp-1\over d_\perp N}}$ \hspace{1.8cm} {\rm relative
acceleration of the branes}\\[2mm]
$n = 1 + 2\eta - 6\epsilon$ & ${\displaystyle
n = 1 - {2(d_\perp-1)\over d_\perp N}}$ 
\hspace{1cm} {\rm fixed by dimensions of compatified space}\\[1mm]
\hline
Number of $e$-folds\\[2mm]
$N = {\displaystyle
{1\over M_{\rm P}}\int{V\,d\phi\over V'(\phi)}}$  &  
$N = {\displaystyle
{H\over v}\int\!dy}$ \hspace{1.9cm} {\rm integrated distance 
between branes}\\[1mm]
\hline
Quantum Fluctuations\\[2mm]
${\cal R}_k = {\displaystyle
{\delta\rho_k\over\dot\rho} = 
H\,{\delta\phi_k\over\dot\phi}}$  &  
${\cal R}_k = {\displaystyle
\delta N_k = {H\over v}\,\delta y_k}$ 
\hspace{9mm} {\rm spatial fluctuations in the branes}\\
\hline
\end{tabular}
\end{table*}

From the point of view of model building this model is very similar to
hybrid inflation; in fact, it is probably the first concrete realization
of hybrid inflation in the context of string
theory~\cite{GBRZ,braneinflation2}.  The main difference is that we do
not have here a complicated field theory with an a priori undetermined
scalar potential; on the contrary, the field content and the string
dynamics is automa\-tically determined by the brane configuration in the
compactified space. This model predicts that the inflaton is massless
while its potential is given by an inverse power law. The usual
slow-roll parameters have here very clear geometrical interpretations as
the relative velocity and acceleration of the branes, and the number of
e-folds is determined by the integrated distance between the branes: the
further apart the branes are initially, the bigger the number of e-folds
of inflation. See Table~\ref{table2} for the general
dictionary between the usual inflationary parameters and the brane
inflation ones. In a sense, the model~\cite{GBRZ} is very robust with
respect to initial conditions and it is not necessary to finetune these
in order to have inflation, see Ref.~\cite{braneinflation}.

The geometrical correspondence goes one step further when studying
quantum fluctuations and the origin of density perturbations. In
ordinary inflation, it is the fluctuations of the inflaton field that
determine the metric fluctuations through the Einstein equations. In
these brane models, inflation ends when the two branes approach
eachother and collide, and the spatial fluctuations of the brane itself
(a sheet in some extra dimensional space) implies that inflation ends at
different times in different points of the (3+1)-brane, directly giving
rise to metric fluctuations. 

The amplitude and tilt of CMB anisotropies in brane inflation models
determines both the string scale and the compactification scale to be of
order $10^{13}$ GeV, slightly lower than usual GUT theories, although
the whole idea of GUT unification has to be revised in the context of
these brane models. Therefore, the model typically predicts a negligible
amplitude for the tensor component in the polari\-zation power spectra
that may escape detection even by Planck. On the other hand, the scalar
tilt is predicted to be $n= 1 - 2(d_\perp-1)/(d_\perp N) = 0.974$, with
$d_\perp=4$ and $N=60$ as in Ref.~\cite{GBRZ}. This is so precise that
the next generation of CMB experiments will very easily rule out this
model in case observations do not agree with this value.

\section{CONCLUSIONS}

After 20+ years, inflation is a robust paradigm with a host of
cosmological observations confirming many of its basic predictions:
large scale spatial flatness and homogeneity, as well as an
approximately scale-invariant Gaussian spectrum of adiabatic density
perturbations.

It is possible that in the near future the next generation of CMB
satellites (MAP and Planck) may detect the tensor or gravitational wave
component of the polarization power spectrum, raising the possibility
of really testing inflation through the comparison of the scalar and
tensor components, as well as determining the energy scale of inflation.

The stage of reheating, and more specifically preheating, after
inflation is today one of the most active areas of research in
theore\-tical cosmology, with the expectation that it may contain clues
to the actual origin of the matter-antimatter asymmetry and be
responsible for signatures, like a stochastic background of
gravitational waves, that may open a new window to the early universe
beyond that of Big Bang Nucleosynthesis.

Although inflation is a very beautiful and elegant paradigm, there is
still no compelling particle physics model for it, while the scale of
inflation is still uncertain. Hybrid inflation is a particularly nice
realization that comes almost for free with any particle physics model
of spontaneous symmetry breaking, be it GUT, intermediate of EW scale,
by assuming a light singlet (the inflaton) coupled to the symmetry
breaking field. 

String theory has recently provided for a concrete realization of
inflation in the context of D-brane interactions mediated by open or
closed strings. It is in fact an interesting hybrid inflation model,
where the string tachyon plays the role of the symmetry breaking field
that triggers the end of inflation. This model is robust and gives
specific predictions for the amplitude and tilt of the spectrum of
metric fluctuations, which will soon be tested against precise
observations of the microwave background anisotropies.

\section*{Acknowledgements}

It is a pleasure to thank the organisers of the XXX Internacional
Meeting on Fundamental Physics, IMFP2002, for their generosity and
enthusiasm in organising a very relaxed and enjoyable meeting. I would
also like to thank David Valls-Gabaud for valuable help with references
on non-Gaussian signatures in the CMB.

\end{document}